\documentclass[
 reprint,
 amsmath,
 amssymb,
 aps,
]{revtex4-1}

\usepackage{latexsym}
\usepackage[usenames]{color}
\usepackage{fancybox}
\usepackage{simplewick}
\usepackage{comment}
\usepackage{framed}
\definecolor{shadecolor}{rgb}{0.9,0.9,0.95}
\usepackage{setspace}
\usepackage{tabularx,booktabs}
\usepackage{bm}
\definecolor{darkgreen}{rgb}{0,0.5,0}
\definecolor{darkblue}{cmyk}{0.9,0.9,0,0}
\definecolor{darkred}{rgb}{0.6,0,0.3}

\usepackage{graphicx}
\usepackage[setpagesize=false,pagebackref=false, linktocpage, bookmarksopen=true, colorlinks=true, linkcolor=darkblue,citecolor=darkblue,urlcolor=darkblue]{hyperref}
\usepackage{hyperref}

\def\del{\partial}

\def\fn#1{\footnote{#1}}
\def\nn{\nonumber}
\def\eqref#1{(\ref{#1})}
\def\comma{\,,}
\def\period{\,.}

\def\beq{\begin{equation}}
\def\eeq{\end{equation}}

\begin{document}
\title{Exact Three-Point Functions of Determinant Operators in Planar $\mathcal{N}=4$ Supersymmetric Yang-Mills Theory}

\author{Yunfeng Jiang$^{1}$, Shota Komatsu$^{2}$, Edoardo Vescovi$^{3}$}

\affiliation{
\vspace{5mm}
$^{1}$Theoretical Physics Department, CERN,  1211 Geneva 23, Switzerland\\
$^{2}$School of Natural Sciences, Institute for Advanced Study, Princeton, New Jersey 08540, USA\\
$^{3}$The Blackett Laboratory, Imperial College, London SW7 2AZ, United Kingdom
}

\begin{abstract}
We introduce a nonperturbative approach to correlation functions of two determinant operators and one non-protected single-trace operator in planar $\mathcal{N}=4$ supersymmetric Yang-Mills theory. Based on the gauge/string duality, we propose that they correspond to overlaps on the string worldsheet between an integrable boundary state and a state dual to the single-trace operator. We determine the boundary state using symmetry and integrability of the dual superstring sigma model, and write down expressions for the correlators at finite coupling, which we conjecture to be valid for operators of arbitrary size. The proposal is put to test at weak coupling.
\end{abstract}

\maketitle
\section{Introduction}
To advance our understanding of nonperturbative dynamics in gauge theories, it is useful to study simple models with rich enough structures. $\mathcal{N}=4$ supersymmetric Yang-Mills theory in four dimensions ($\mathcal{N}=4$ SYM) is one of the leading candidates for the following reasons: First it admits the planar large $N_c$ limit, which makes it amenable to analytical studies. Second it is a conformal field theory (CFT) and all the correlation functions can be decomposed into two- and three-point functions. Third it can be described alternatively in terms of two-dimensional string worldsheets which we can analyze exactly using integrability.
The application of integrability led to a complete determination of two-point functions of local operators \cite{Beisert:2010jr}.
It was applied also to the three-point function \cite{Basso:2015zoa}, but the result is still unsatisfactory since it is given by a series expansion which one needs to resum.

In this letter, we present the first fully nonperturbative result for the three-point function valid for a large class of operators \fn{Some three-point functions are already known at finite coupling. For instance, the structure constant of a Lagrangian operator and two identical operators is given by a derivative of the conformal dimension $\del_{\lambda}\Delta$ \cite{Costa:2010rz}. However, it is determined solely by the conformal dimension and does not provide truly new conformal data.}. Specifically, we study the correlator of two determinant operators and one non-protected single-trace operator. By interpreting this correlator as an overlap on the string worldsheet between a boundary state and a state dual to the single-trace operator, we write down nonperturbative expressions using the framework of thermodynamic Bethe ansatz (TBA) \cite{Zamolodchikov:1991et}.
\vspace*{-0.7 cm}
\section{Set Up and Basic Strategy}
The main subject is the three-point function of a non-protected single-trace operator $\mathcal{O}$ and two determinant operators $\mathcal{D}_{1,2}\equiv \det \mathfrak{Z}(a_{1,2})$ \footnote{We chose a configuration suitable for analyzing the symmetry. More general configurations can be obtained by performing the conformal and the R-symmetry transformations. The structure constant $\mathfrak{D}_{\mathcal{O}}$ is not affected by such transformations.} with
\begin{align}
\mathfrak{Z}(a)\equiv \left.\frac{(1+a^2)\Phi^1+i(1-a^2)\Phi^2+2i a\Phi^{4}}{\sqrt{2}}\right|_{\substack{x^{\mu}=\quad\\(0,a,0,0)}}\,,
\end{align}
where $\Phi^{1,2,4}$ are real scalar fields in $\mathcal{N}=4$ SYM. Owing to the superconformal symmetry, its spacetime dependence is fixed to be  \cite{Basso:2015zoa,Drukker:2009sf}
\begin{align}\label{eq:3pt}
\langle \mathcal{D}_1 \mathcal{D}_2 \mathcal{O}(0)\rangle=\left(\frac{a_1-a_2}{a_1 a_2}\right)^{\Delta-J}\mathfrak{D}_{\mathcal{O}}\,,
\end{align}
where $\mathfrak{D}_{\mathcal{O}}$ is the structure constant while $\Delta$ and $J$ are the conformal dimension and the R-charge of $\mathcal{O}$.

\begin{figure}[t]
\centering
\includegraphics[scale=0.11]{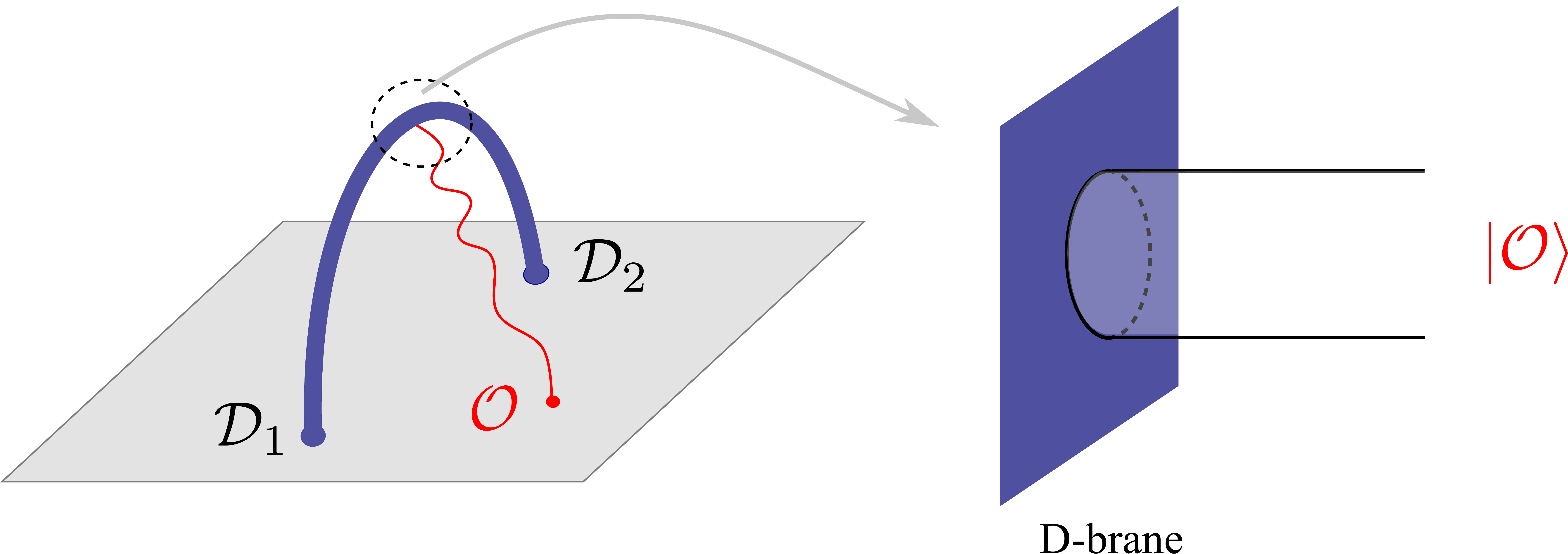}
\caption{The AdS description of the three-point function. The thick curve represents a geodesic of the D-brane dual to the determinant operators while the wavy line denotes a closed string dual to the single trace operator. On the worldsheet, it corresponds to an overlap with a boundary state.}
\label{fig:geodesic}
\end{figure}

\begin{figure}[h!]
\centering
\includegraphics[scale=0.1]{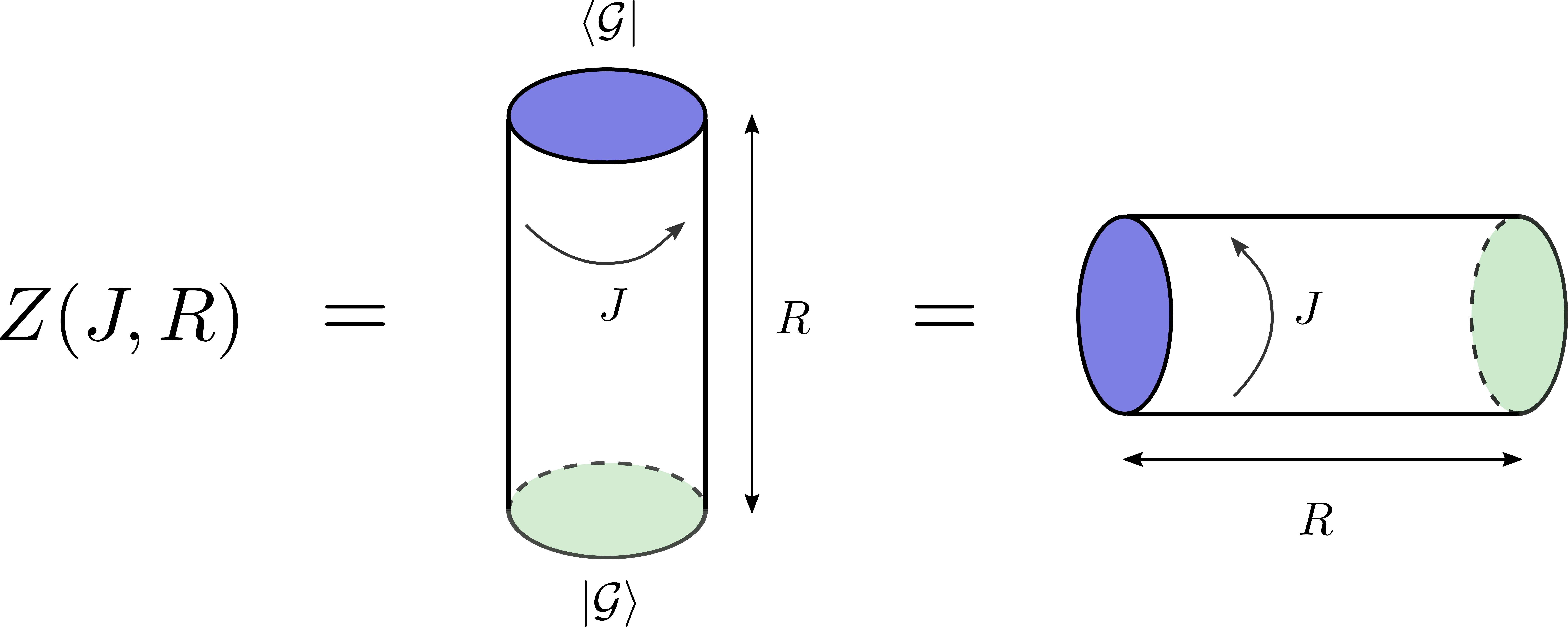}
\caption{The partition function $Z(J,R)$ evaluated in two different channels. The left/right panel denote the closed-/open- string channel of the same partition function. To compute the overlap $\langle \mathcal{G}|\Omega\rangle$, we take the limit where $J$ is finite and $R\to\infty$.}
\label{fig:channels}
\end{figure}

The goal of this letter is to compute  $\mathfrak{D}_{\mathcal{O}}$ nonperturbatively using the gauge/string duality.
As discussed in \cite{Bissi:2011dc,Balasubramanian:2001nh,McGreevy:2000cw}, the duality maps \eqref{eq:3pt} to a closed string in AdS$_5\times$S$^5$ which ends on a geodesic of a D-brane dual to determinant operators (see Figure \ref{fig:geodesic}).
On the string worldsheet, this corresponds to an overlap between a boundary state $\langle \mathcal{G}|$ and a state dual to $\mathcal{O}$. To evaluate such an overlap, we first consider a partition function $Z(J,R)$ of a cylinder worldsheet whose ends are capped off by the boundary states (see Figure~\ref{fig:channels}).
In the limit $R\to \infty$, the expansion of $Z(J,R)$ in the closed string channel is dominated by the ground state $|\Omega\rangle$
\begin{align}
\begin{aligned}\label{eq:partitionFunction}
Z(J,R)&=\sum_{\psi_{\rm c}} \langle \mathcal{G}|\psi_{\rm c}\rangle e^{-E_{\psi_{\rm c}}R}\langle\psi_{\rm c} |\mathcal{G}\rangle\\&\overset{R\to \infty}{\to} |\langle \mathcal{G}|\Omega\rangle|^2 e^{-E_{\Omega}R}\period
\end{aligned}
\end{align}
By contrast, in the open string channel, $Z(J,R)$ can be viewed as the thermal free energy, and the limit corresponds to the thermodynamic limit in which the volume of the space becomes infinite. This allows us to compute $\langle \mathcal{G}|\Omega\rangle$ using TBA. The result for excited states can be obtained from $\langle \mathcal{G}|\Omega\rangle$ by analytic continuation \cite{Dorey:1996re}.
\vspace*{-0.3 cm}
\section{Constraints on Boundary States}\label{sec:constraints}
To apply the aforementioned strategy, we first determine the boundary state $\langle \mathcal{G}|$ in the infinite volume ($J\to \infty$) limit. For this, we assume that $\langle \mathcal{G}|$ is an {\it integrable boundary state}, namely a state corresponding to a boundary condition which preserves infinite many conserved charges \cite{Ghoshal:1993tm}. The assumption is justified a posteriori by agreement with weak-coupling computations as we see later. For integrable boundary states, the overlap in the $J\to\infty$ limit  can be factorized into two-particle overlaps
\begin{align}
F_{{\bf A}{\bf B}}(u)\equiv \left.\langle \mathcal{G}|\mathcal{X}_{\bf A}(u)\mathcal{X}_{\bf B}(\bar{u})\rangle\right|_{J\to \infty}\,,
\end{align}
where $\mathcal{X}$'s are magnons in the $\mathcal{N}=4$ SYM spin chain, and ${\bf A}=A\dot{A}$ and ${\bf B}=B\dot{B}$ are in the bifundamental representation of the $\mathfrak{psu}(2|2)^2$ symmetry \cite{Beisert:2005tm}. The rapidities $u$ and $\bar{u}$ are parity-conjugate to each other, and satisfy
$
x^{\pm}(\bar{u})=-x^{\mp}(u)$
where $f^{\pm} (u)\equiv f(u\pm \frac{i}{2})$ and the Zhukovsky variable $x(u)$ is defined by
$
x(u)\equiv \frac{u+\sqrt{u^2-4g^2}}{2g}
$ with $g\equiv \frac{\sqrt{\lambda}}{4\pi}$ and $\lambda$ being the 't Hooft coupling constant.

\begin{figure}[h!]
\centering
\includegraphics[scale=0.13]{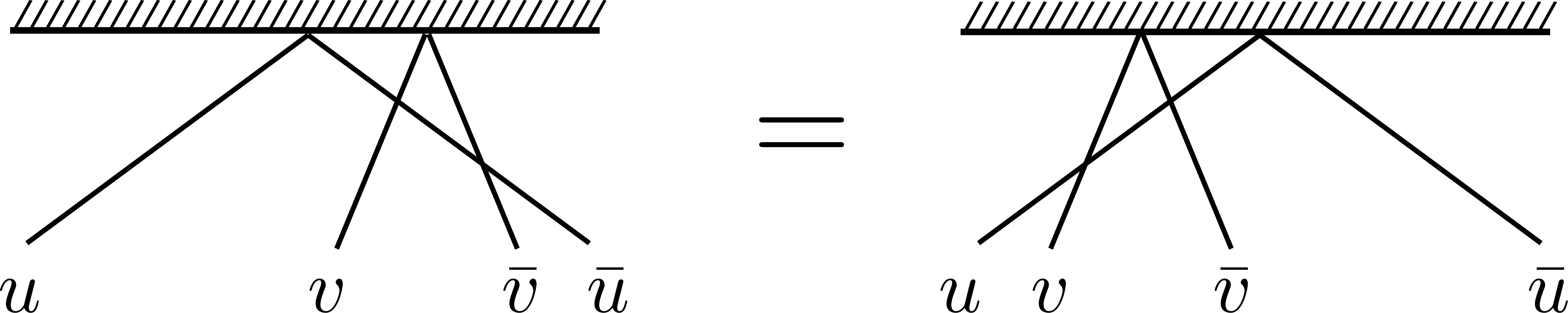}
\caption{The boundary Yang-Baxter equation.}
\label{fig:bYBE}
\end{figure}

\begin{figure}[h!]
\centering
\includegraphics[scale=0.12]{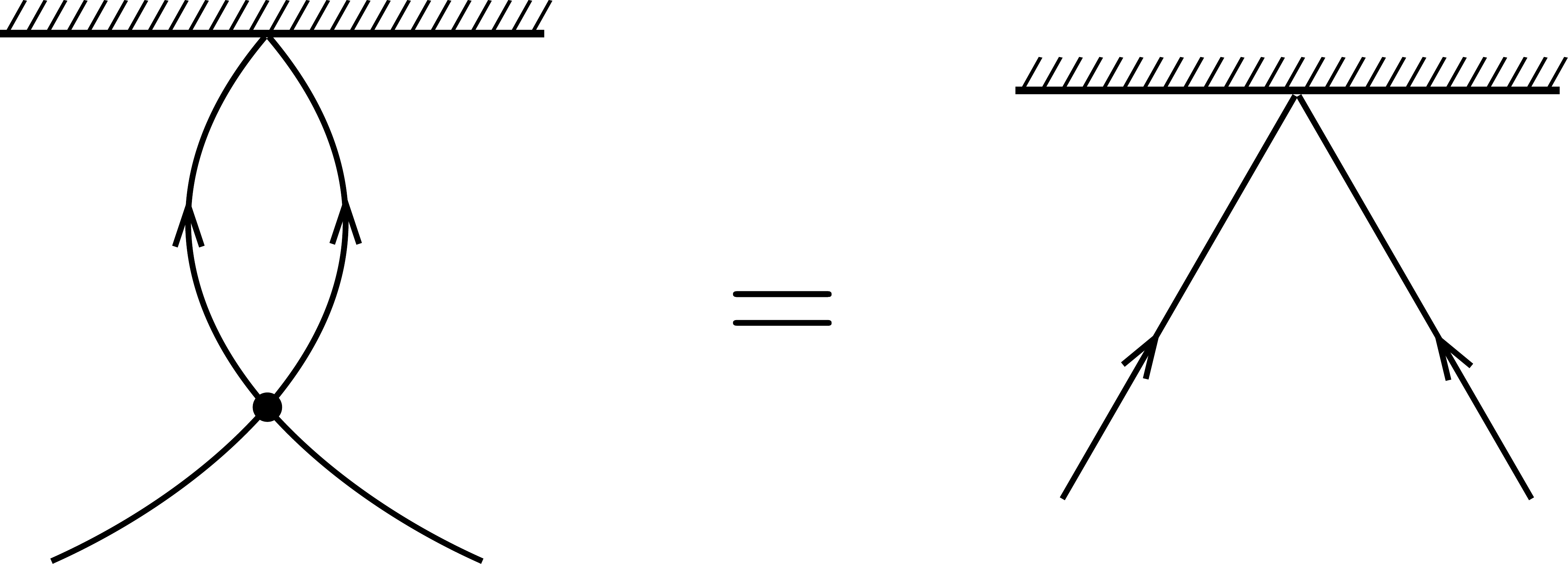}
\caption{Watson's equation for the two-particle overlap.}
\label{fig:axioms1}
\end{figure}

\begin{figure}[h!]
\centering
\includegraphics[scale=0.12]{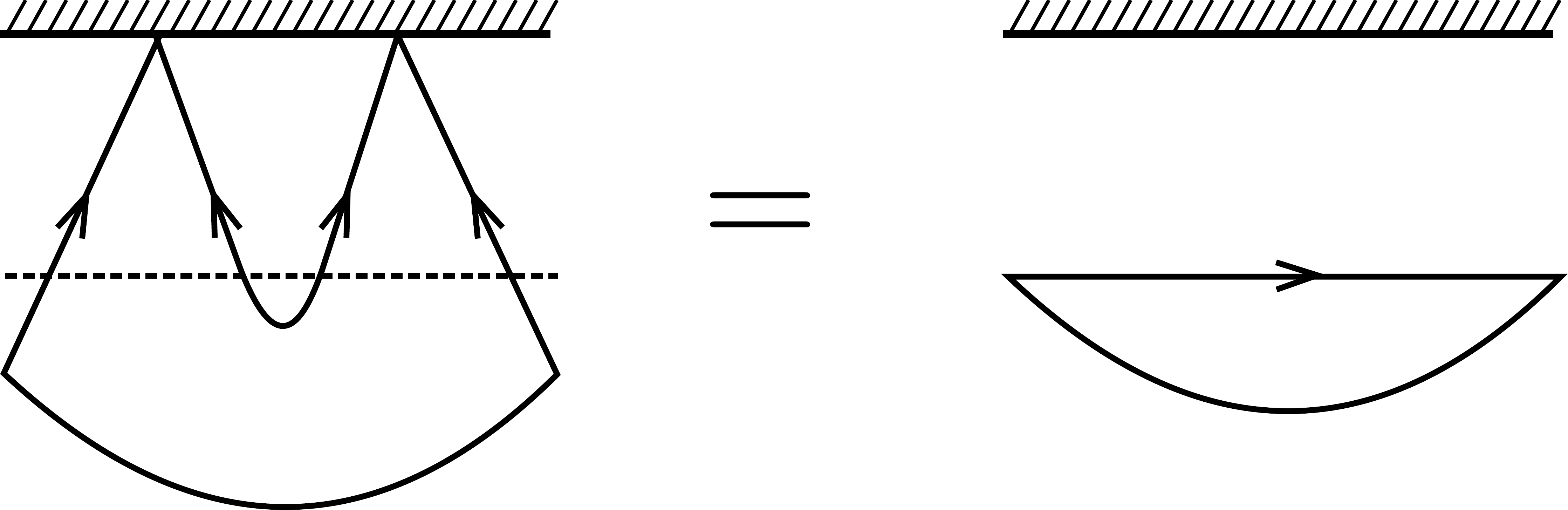}
\caption{Decoupling equation for the two-particle overlap.}
\label{fig:axioms2}
\end{figure}

\paragraph{Boundary Yang-Baxter equation}
The integrable boundary state satisfies the so-called boundary Yang-Baxter equation (bYBE), which reads (see Figure~\ref{fig:bYBE})
\begin{align}\label{eq:bYBE}
\begin{aligned}
&\left.\langle \mathcal{G}|\mathbb{S}_{24}\mathbb{S}_{34}|\mathcal{X}_{1}(u)\mathcal{X}_{2}(v)\mathcal{X}_{3}(\bar{v})\mathcal{X}_{4}(\bar{u})\rangle\right|_{J\to \infty}\\
=&\left.\langle \mathcal{G}|\mathbb{S}_{13}\mathbb{S}_{12}|\mathcal{X}_{1}(u)\mathcal{X}_{2}(v)\mathcal{X}_{3}(\bar{v})\mathcal{X}_{4}(\bar{u})\rangle\right|_{J\to \infty}\comma
\end{aligned}
\end{align}
where $\mathbb{S}_{kl}$ is the bulk $S$-matrix \cite{Beisert:2005tm} between $\mathcal{X}_k$ and $\mathcal{X}_l$.
\paragraph{Watson's equation}
The second constraint is Watson's equation, which states that an exchange of particles is equivalent to a multiplication of the S-matrix. Explicitly, it reads (see Figure \ref{fig:axioms1})
\begin{align}
\label{eq:watson1}
F_{\mathbf{A}\mathbf{B}}(u)=\mathbb{S}_{\mathbf{A}\mathbf{B}}^{\mathbf{C}\mathbf{D}}(u,\bar{u})F_{\mathbf{C}\mathbf{D}}(\bar{u})\,.
\end{align}
\vspace{-0.5cm}
\paragraph{Decoupling equation}
The last condition is the decoupling condition, which is equivalent to the boundary unitarity in \cite{Ghoshal:1993tm}. It states that a pair of particle-antiparticle pairs must decouple from the rest of the overlap (see Figure \ref{fig:axioms2}) and reads
\begin{align}
F_{{\bf A}{\bf B}}(u)
\mathfrak{C}^{{\bf B}{\bf B}^{\prime}}F_{{\bf B}^{\prime}{\bf C}^{\prime}}(\bar{u}^{2\gamma})\mathfrak{C}^{{\bf C}^{\prime}{\bf C}}=\delta_{\bf A}^{\bf C}
\end{align}
where $\mathfrak{C}$ is the charge conjugation matrix \cite{Janik:2006dc}, and $ u^{2\gamma}$ is the crossing transformation defined by
$
x^{\pm}(u^{2\gamma})=\frac{1}{x^{\pm}(u)}$.
\paragraph{Solution} Solving these constraints, the two-particle overlap is fixed to be
\beq\label{eq:twoparticleexplicit}
F_{{\bf A}{\bf B}}(u)=\frac{x^{+}}{x^{-}}\frac{u-\frac{i}{2}}{u}\frac{\sigma_B (u)}{\sigma (u,\bar{u})}(-1)^{|\dot{A}||B|}M_{{\bf A},{\bf B}}\comma
\eeq
where $|\bullet|$ denotes the grading of the index $\bullet$ and $\sigma (u,v)$ is the bulk dressing phase \cite{Beisert:2006ez}. As shown in Appendix \ref{ap:reflection}, there are two choices for the matrix part $M_{{\bf A},{\bf B}}$ and we conjecture that the true boundary state is given by a sum of the two. See also discussions below \eqref{eq:finalSL2finite}. $\sigma_B(u)$ is the {\it boundary dressing phase} satisfying \beq
\sigma_B (\bar{u})=\sigma_B(u)\comma\qquad \sigma_B (u)\sigma_B (u^{2\gamma})=\frac{u^2}{u^{2}+\frac{1}{4}}\period
\eeq
A solution is given by $\sigma_B (u)=4^{\frac{1+x^{+}x^{-}}{1-x^{+}x^{-}}}\frac{G(x^{+})}{G(x^{-})}$
with
\begin{align}
\begin{aligned}
\log G(x)=&\,\oint \frac{dz}{2\pi i}\frac{\log \mathfrak{G}(g(z+z^{-1}))}{x-z}\comma
\end{aligned}
\end{align}
and $\mathfrak{G}(u)\equiv\,\frac{\Gamma \left(\frac{1}{2}-iu\right)\Gamma \left(1+iu\right)}{\Gamma \left(\frac{1}{2}+iu\right)\Gamma \left(1-iu\right)}$.
\section{$g$-Function for Ground State}
\begin{figure}[t]
\centering
\includegraphics[scale=0.10]{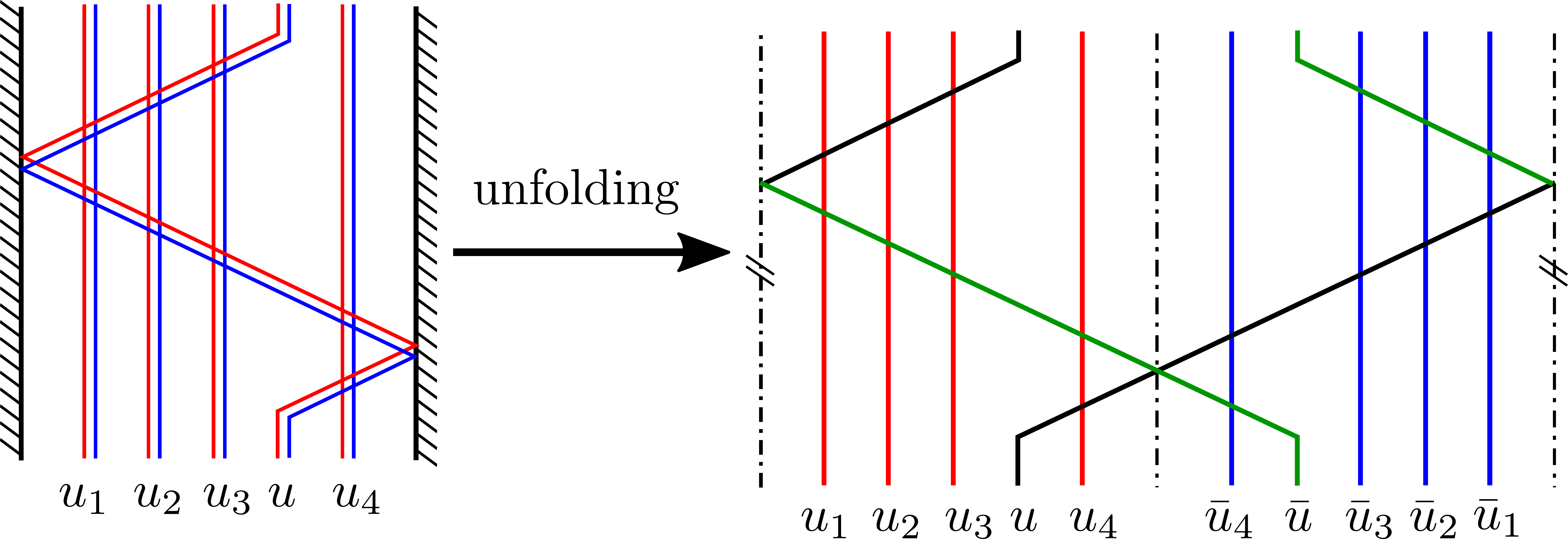}
\caption{The bABA and its unfolding interpretation. The structure of the reflection matrix allows us to map the bABA to an ABA of a closed chain with a single $\mathfrak{psu}(2|2)$ symmetry.}
\label{fig:unfolding}
\end{figure}

We now discuss the ground-state overlap $\langle \mathcal{G}|\Omega\rangle$ for finite $J$.
For this we consider $Z(J,R)$ in the open string channel (also known as the {\it mirror channel}) and take the limit $R\to \infty$:
\beq\label{eq:pathintegralrho}
Z(J,R)=\sum_{\psi_{o}}e^{-\tilde{E}_{\psi_{o}}J} \,\overset{R\to \infty}{\to}\, \mathcal{N}\int \mathcal{D}\rho \,e^{-RS_{\rm eff}[\rho]}\period
\eeq
As shown above, in the limit $R\to \infty$ one can replace the sum over $\psi_{o}$ with a path integral of densities $\rho$.
\paragraph{Bethe equation in the mirror channel} The crucial input for writing down $S_{\rm eff}$ is the boundary asymptotic Bethe equation (bABA), which constrains the rapidities of magnons. Schematically, it reads (see Figure \ref{fig:unfolding})
\beq\label{eq:bABA}
{\bf 1}=\mathbb{R}_{L}(u_j)\prod_{k\neq j}\mathbb{S}(u_j,u_k)\,\,\mathbb{R}_R (u_j)\,\,\prod_{k\neq j}\mathbb{S}(u_j,\bar{u}_k)\comma
\eeq
where $\mathbb{R}_{L,R}$ are the left/right reflection matrices. The reflection matrices are related to the infinite-volume overlap \eqref{eq:twoparticleexplicit} by $[\mathbb{R}_{L}]_{{\bf A}}^{{\bf B}}(u)=[\mathbb{R}_{R}]_{{\bf A}}^{{\bf B}}(\bar{u})=F_{{\bf A}{\bf  C}}(u^{\gamma})\mathfrak{C}^{{\bf C}{\bf B}}$ with $u^{\gamma}$ being the mirror transformation defined by
$
x^{+}(u^{\gamma})=1/x^{+}(u)$ and $x^{-}(u^{\gamma})=x^{-}(u)
$.
As a result, we find
\begin{align}
\begin{aligned}
\left[\mathbb{R}_{L}\right]^{B\dot{B}}_{A\dot{A}}(u)=\frac{u-\frac{i}{2}}{u}\frac{\sigma_B (u^{\gamma})}{\sigma (\bar{u}^{\gamma},u^{\gamma})}\mathcal{S}_{A\dot{A}}^{B\dot{B}}(\bar{u}^{\gamma},u^{\gamma})\comma
\end{aligned}
\end{align}
where $\mathcal{S}$ is a single copy of the $\mathfrak{psu}(2|2)$ S-matrix \cite{Beisert:2005tm}. The structure of $\mathbb{R}_{L,R}$ allows for the {\it unfolding} interpretation; the bABA \eqref{eq:bABA} can be mapped to an ABA of a closed string with a single $\mathfrak{psu}(2|2)$ symmetry (see Figure \ref{fig:unfolding}).

\paragraph{TBA equation} $S_{\rm eff}$ can be derived from ABA following the standard derivation of TBA. In the $R\to \infty$ limit, \eqref{eq:pathintegralrho} can be approximated by the saddle point $\frac{\delta S_{\rm eff}}{\delta \rho}=0$. Due to the unfolding structure, the saddle-point equations coincide with the standard TBA for the spectrum \cite{Gromov:2009tv,Bombardelli:2009ns,Arutyunov:2009ur} with the identification $Y_{a,s}(\bar{u})=Y_{a,-s}(u)$. They take a form of $\log Y_{a,s}=\varphi^{+}_{a,s}$. For instance, $\varphi^{+}_{a,0}$ reads (the full equations are given in \cite{Jiang:2019xdz})
\begin{align}\label{eq:TBAphase}
&\varphi^{+}_{a,0}\equiv \,-J\tilde{E}_a+\log(1+Y_{b,0})\ast (\mathcal{K}_+)^{\bullet\bullet}_{b,a}
\nn\\
&+\log(1+Y_{m,1})\star(\mathcal{K}_+)^{\triangleright\bullet}_{m-1,a}+\log(1+1/Y_{2,2})\,\,\widehat{\star}\,\,(\mathcal{K}_+)^{y\bullet }_{+a}\nn\\
&+\log(1+Y_{1,1})\,\,\widehat{\star}\,\,(\mathcal{K}_+)^{y\bullet }_{-a}\,.
\end{align}
Here we follow the notations in \cite{Bajnok:2010ke}, and $\star$, $\ast$, and $\widehat{\star}$ denote the convolutions along $[-\infty,\infty]$, $[0,\infty]$ and $[-2g,2g]$ respectively. $\mathcal{K}_{+}$ is a symmetrized kernel defined by $K(u,v)+K(u,\bar{v})$ with $K$ being the standard TBA kernel.
\paragraph{g-function} The saddle point value of $S_{\rm eff}$ only gives the exponentially decaying piece in \eqref{eq:partitionFunction}, $e^{-E_{\Omega}R}$. To read off the overlap, we need to consider the one-loop fluctuation around the saddle point and the $O(1)$ normalization factor $\mathcal{N}$ in \eqref{eq:pathintegralrho}. Such analysis was performed in the literature \cite{Dorey:2009vg,Pozsgay:2010tv,Kostov:2018dmi}, and the application to our problem leads to an expression for the ground states, which corresponds to BPS operators in $\mathcal{N}=4$ SYM,

\begin{align}\label{eq:gfiniteGandGplus}
\begin{aligned}
\langle \mathcal{G}|\Omega\rangle &=e^{\sum_{a}\int_0^{\infty}\frac{du}{2\pi}\Theta_a \log (1+Y_{a,0})}
 \frac{\sqrt{{\rm Det}(1-\hat{G})}}{{\rm Det} (1-\hat{G}_{+})}\,.
\end{aligned}
\end{align}
Here $a$ runs from $1$ to $\infty$ and $
\Theta_a (u)=i\del_u \log r_a (\bar{u})-\pi \delta (u)+\frac{i}{2}\del_u \log S_{aa}^{\bullet\bullet} (u,\bar{u})
$ where $S_{ab}^{\bullet\bullet}$ is the bound-state S-matrix and $r_a$ is the bound-state reflection factor given in Appendix \ref{ap:reflection}. ${\rm Det}$ denotes the Fredholm determinant \footnote{In what follows, ${\rm Det}$ denotes the Fredholm determinant while ${\rm det}$ denotes a standard determinant of a finite-dimensional matrix.} and $\hat{G}_{+}$ is an integral kernel defined by $[\hat{G}_{+}]^{(a,s)}_{(b,t)}(u,v)\equiv \frac{\delta \varphi^{+}_{a,s}(u)}{\delta \log Y_{b,t}(v)}$. Similarly, $\hat{G}$ is given by $\frac{\delta \varphi_{a,s}(u)}{\delta \log Y_{b,t}(v)}$ where $\varphi_{a,s}$ are the ``right hand sides'' of the standard TBA, $\log Y_{a,s}=\varphi_{a,s}$, without identification of $Y$-functions.
\section{Conjecture for SL(2) Sector}
We now generalize \eqref{eq:gfiniteGandGplus} to excited states in the SL(2) sector using the analytic continuation trick \cite{Dorey:1996re} following the standard TBA analysis.
\paragraph{$g$-function for excited states}
After the analytic continuation, poles of $1/(1+1/Y_{1,0})$ cross the integration contour and modify the overlap \eqref{eq:gfiniteGandGplus}.
As a result, we find that the structure constant $\mathfrak{D}_{\mathcal{O}}$ is given by
\beq\label{eq:finalSL2finite}
\begin{aligned}
\mathfrak{D}_{\mathcal{O}}
&=-\frac{i^{J}+(-i)^{J}}{\sqrt{J}}\,e^{\sum_{a}\int_0^{\infty}\frac{du}{2\pi}\Theta_a \log (1+Y_{a,0})}\\
&\times\sqrt{\prod_{1\leq s\leq \frac{M}{2}}\frac{u_s^2+\frac{1}{4}}{u_s^2}\sigma_B^2(u_s)}\frac{\sqrt{{\rm Det} (1-\hat{G}^{\bullet})}}{{\rm Det} (1-\hat{G}^{\bullet}_{+})}\period
\end{aligned}
\eeq
This is the main result of this letter which we conjecture to be valid for any length $J$ and at finite $\lambda$.
Here the $J$-dependent prefactor reflects the fact that the true boundary state is a sum of two boundary states as mentioned below \eqref{eq:twoparticleexplicit}. $\hat{G}^{\bullet}$ and $\hat{G}_{+}^{\bullet}$ are given by \footnote{The kernels in this letter are related to the ones in \cite{Jiang:2019xdz} by the transposition.}
\begin{align}\label{eq:modifiedkernel}
\begin{aligned}
\hat{G}^{\bullet}\cdot f=\sum_{k=1}^{M}\frac{iK^{\bullet, X}_{1,x}(u_k,u)}{\partial_u\log Y_{1,0}(u_k)}f(u_k)
+\hat{G}\cdot f\comma\\
\hat{G}_{+}^{\bullet}\cdot f=\sum_{k=1}^{M/2}\frac{i(\mathcal{K}_{+})^{\bullet, X}_{1,x}(u_k,u)}{\partial_u\log Y_{1,0}(u_k)}f(u_k)
+\hat{G}_{+}\cdot f\period
\end{aligned}
\end{align}
Here and below, $x$ and $X$ take various indices and symbols which represent different bound states. The sum in \eqref{eq:modifiedkernel} come from the poles crossing the contours, and $u_k$'s are the magnon rapidities satisfying the parity condition,
\beq \label{eq:paritycond}
u_{\frac{M}{2}+k}=\bar{u}_k \qquad (1\leq k\leq M/2)\period
\eeq
They are the solutions to the {\it exact Bethe equations} \cite{Gromov:2009tv},
\beq\label{eq:exactBethe}
\phi (u_j)=2\pi i\left(n_j+\frac{1}{2}\right)\comma\qquad n_j \in \mathbb{Z}\comma
\eeq
with
\beq\label{eq:defphi}
\begin{aligned}
\phi (u)\equiv&  \,-J\tilde{E}_1+\sum_{k=1}^{M} \log S^{\bullet \bullet}_{11} (u,u_k)\\
&+\sum_{x,X}\log (1+Y_{X,x})\star K^{X,\bullet}_{x,1}\period
\end{aligned}
\eeq
\paragraph{Exact Gaudin determinants} The result \eqref{eq:modifiedkernel} can be rewritten into a form similar to the so-called Gaudin determinants. For this, we first split the kernel $\hat{G}^{\bullet}$ into a sum ${\sf S}$ and an integral ${\sf I}$ (see \eqref{eq:modifiedkernel}), and rewrite ${\rm Det}(1-\hat{G}^{\bullet})$ as ${\rm Det}(1-{\sf S}-{\sf I})={\rm Det}(1-\tilde{\sf S})\times {\rm Det}(1-{\sf I})$ with $\tilde{\sf S}\equiv {\sf S}/(1-{\sf I})$. Similarly, ${\rm Det}(1-\hat{G}^{\bullet}_{+})$ can be split into a sum ${\sf S}_{+}$ and an integral ${\sf I}_{+}$, and can be re-expressed as ${\rm Det}(1-\hat{G}^{\bullet}_{+})={\rm Det}(1-\tilde{\sf S}_{+})\times {\rm Det}(1-{\sf I}_{+})$ with $\tilde{\sf S}_{+}\equiv {\sf S}_{+}/(1-{\sf I}_{+})$.

Next we consider $\del_{u_k} \phi (u_j)$ $(j,k=1,\ldots, M)$. The derivative $\del_{u_k}$ can act either on $p(u_k)$, $\log S^{\bullet\bullet}_{11}$ or $Y_{X,x}$ in \eqref{eq:defphi}. We then eliminate $\del_{u_k} Y_{X,x}$ by considering the {\it excited state TBA} (see \cite{Gromov:2009tv} for the full set of equations)
\beq
\begin{aligned}
\log Y_{a,0}=&-J \tilde{E}_a+\sum_{k=1}^{M}\log S^{\bullet\bullet}_{a,1}(u,u_k)\\
&+\sum_{x,X}\log (1+Y_{X,x})\star K_{x,a}^{X,\bullet}\comma
\end{aligned}
\eeq
taking a derivative $\del_{u_k}$ of both sides, and solving for $\del_{u_k}Y_{X,x}$. The parity condition \eqref{eq:paritycond} is imposed only at the end of the computation. As a result of these manipulations, we find that $\det (\del_{u_k}\phi(u_j))\propto {\rm Det}(1-\tilde{\sf S})$ up to some constant of proportionality. The relation in particular shows that ${\rm Det}(1-\tilde{\sf S})$ is actually a finite-dimensional determinant although it was initially defined as the Fredholm determinant. Details of the rewriting are explained in a toy example in Appendix \ref{ap:rewriting}.

On the other hand, if we first impose the parity condition \eqref{eq:paritycond} and compute the derivatives $\del_{u_k}\phi(u_j)$, we find that $\det (\del_{u_k}\phi(u_j))$ $(j,k=1,\ldots \frac{M}{2})$ is now proportional to ${\rm Det}(1-\tilde{\sf S}_{+})$. Upon taking the ratio, the constants of proportionality cancel out and we obtain
\beq\label{eq:ratiotoexactG}
\frac{\sqrt{{\rm Det}(1-\tilde{\sf S})}}{{\rm Det}(1-\tilde{\sf S}_{+})}=\frac{\sqrt{\det (\del_{u_k}\phi(u_j))}}{\det (\del_{u_k}\phi^{+}(u_j))}\comma
\eeq
where $\phi^{+}$ denotes that we are imposing the parity condition before computing derivatives. These determinants can be viewed as the finite-volume version of the Gaudin determinants for the norm of the spin chain. They also resemble the finite-volume one-point functions in sin(h)-Gordon model \cite{Jimbo:2010jv,Negro:2013wga,Bajnok:2019yik}.

Using this rewriting, we obtain an alternative representation for the Fredholm determinants in \eqref{eq:finalSL2finite},
\beq\label{eq:newfredholm}
\frac{\sqrt{{\rm Det} (1-\hat{G}^{\bullet})}}{{\rm Det} (1-\hat{G}^{\bullet}_{+})}=\frac{\sqrt{\det (\del_{u_k}\phi(u_j))}}{\det (\del_{u_k}\phi^{+}(u_j))}\frac{\sqrt{{\rm Det}(1-\tilde{\sf I})}}{{\rm Det}(1-\tilde{\sf I}_{+})}\period
\eeq
\paragraph{Asymptotic formula} Using the representation \eqref{eq:newfredholm}, one can take the {\it asymptotic limit} of \eqref{eq:finalSL2finite},  in which the size of the operator becomes large $J\gg 1$. In this limit, the middle-node Y-functions are exponentially suppressed, $Y_{a,0}\to 0$, and one can show that both
\beq
e^{\sum_{a}\int_0^{\infty}\frac{du}{2\pi}\Theta_a \log (1+Y_{a,0})}\quad \text{and}\quad \frac{\sqrt{{\rm Det}(1-\tilde{\sf I})}}{{\rm Det}(1-\tilde{\sf I}_{+})}\nonumber
\eeq
tend to unity. We thus obtain the following expression for the structure constant in the asymptotic limit:
\beq\label{eq:asymptformula}
\begin{aligned}
\mathfrak{D}_{\mathcal{O}}^{\rm asym}=&-\frac{i^{J}+(-i)^{J}}{\sqrt{J}}\sqrt{\prod_{1\leq s\leq \frac{M}{2}}\frac{u_s^2+\frac{1}{4}}{u_s^2}\sigma_B^2(u_s)}\\
&\times \frac{\sqrt{\det (\del_{u_k}\phi(u_j))}}{\det (\del_{u_k}\phi^{+}(u_j))}\period
\end{aligned}
\eeq
Note that the determinants on the second line are the standard Gaudin-like determinants since all the finite-size corrections can be dropped.
For generalization of \eqref{eq:asymptformula} to operators outside the SL(2) sector, see \cite{Jiang:2019xdz}. A similar formula was found at weak coupling for the defect one-point functions \cite{deLeeuw:2015hxa}.
\vspace*{-0.2 cm}
\section{Weak Coupling Test}
To test our formula \eqref{eq:asymptformula}, we computed the four-point function of $\mathcal{D}_{1,2}$ and two ${\bf 20}^{\prime}$ operators $\mathcal{O}_{{\bf 20}^{\prime}}$ up to $O(\lambda^2)$. We then performed the operator product expansion to read off the conformal data of the spin-$S$ twist-2 operators $\mathcal{O}_S$. The details are given in \cite{Jiang:2019xdz}.

The results of the computation are summarized in Table \ref{tab:squareOPE}. We compared them against the integrability prediction \eqref{eq:asymptformula} and observed a perfect match. This is quite a nontrivial test of our formalism since the results contain the transcendental number $\zeta_3$ and include the contributions from the boundary dressing phase $\sigma_B(u)$. Further tests at weak and strong couplings are provided in \cite{Jiang:2019xdz}.
\begin{table}[t]
\caption{The squared structure constants $\left(\mathfrak{D}_{\mathcal{O}_{S}}\right)^{2}$ for the spin-$S$ twist-2 operators.}
\centering
\begin{tabular}{l|l}
$S$&\multicolumn{1}{c}{$\left(\mathfrak{D}_{\mathcal{O}_{S}}\right)^{2}$}\\
\toprule
$2$&$\frac{1}{3}-4 g^2+ g^4(56-24\zeta_3)$\\
$4$&$\frac{1}{35}-\frac{205 g^2}{441}+g^4\left(\frac{70219}{9261}-\frac{20\zeta_3}{7}\right)$\\
$6$&$\frac{1}{462}-\frac{1106
   g^2}{27225}+g^4\left(\frac{772465873}{1078110000}-\frac{14\zeta_3}{55}\right)$\\
$8$&$\frac{1}{6435}-\frac{14380057
   g^2}{4509004500}+g^4\left(\frac{5048546158688587}{85305405235050000}-\frac{1522\zeta_3}{75075}\right)$
\end{tabular}\label{tab:squareOPE}
\end{table}

We also found that the structure constants exhibit a simple large spin behavior up to two loops,
\beq
\begin{aligned}
\log \left[\frac{\mathfrak{D}_{\mathcal{O}_{S}}}{\left.\mathfrak{D}_{\mathcal{O}_{S}}\right|_{\text{tree level}}}\right]=&f_1 \log S^{\prime}+f_2+O(1/S^{\prime})
\end{aligned}
\eeq
with $\log S^{\prime}\equiv \log S +\gamma_E$, where $\gamma_E$ is the Euler-Mascheroni constant, and
\beq
\begin{aligned}
f_1 &=-4 g^2 \log 2+8g^4\left[\zeta_2\log 2+\frac{9}{2}\zeta_3\right]+O(g^6)\comma\\
f_2&=-2g^2\zeta_2+8g^4\left[\frac{4}{5}(\zeta_2)^2+\frac{3}{2}\zeta_3\log 2\right]+O(g^6)\period
\end{aligned}
\eeq
\section{Conclusion}
In this letter, we applied the TBA formalism to write down a nonperturbative expression for the structure constant of two determinant operators and a single-trace operator in the SL(2) sector of arbitrary size. Our result would provide a foundation for future developments, such as the reformulation in terms of the Quantum Spectral Curve \cite{Gromov:2013pga}, as was the case with the TBA for the spectrum. It would also be worth trying to extract various interesting physics from our formula. We also hope that our approach gives useful insights into the three-point functions of single-trace operators \cite{Basso:2015zoa}.
\paragraph*{Acknowledgement}
We thank Amit Sever for helpful discussions. SK is supported by DOE
grant number DE-SC0009988. EV is funded by the FAPESP grants 2014/18634-9 and 2016/09266-1, and by
the STFC grant ST/P000762/1.
\bibliography{LetterRef}

\newpage
\appendix
\section{Overlap and Reflection Matrix}\label{ap:reflection}
The matrix part of the two-particle overlap is given by
\beq\label{eq:matrixpartwithz}
\begin{aligned}
M_{a\dot{a},b\dot{b}}&=\frac{1+B}{2}\epsilon^{a\dot{b}}\epsilon^{b\dot{a}}+\frac{1-B}{2}\epsilon^{a\dot{a}}\epsilon^{b\dot{b}}\comma\\
M_{\alpha\dot{\alpha},\beta\dot{\beta}}&=\frac{-1+E}{2}\epsilon^{\alpha\dot{\beta}}\epsilon^{\beta\dot{\alpha}}-\frac{1+E}{2}\epsilon^{\alpha\dot{\alpha}}\epsilon^{\beta\dot{\beta}}\comma\\
M_{a\dot{a},\beta\dot{\beta}}&=G\epsilon^{a\dot{a}}\epsilon^{\beta\dot{\beta}}\comma\quad
M_{\alpha\dot{\alpha},b\dot{b}}=G\epsilon^{\alpha\dot{\alpha}}\epsilon^{b\dot{b}}\comma\\
M_{a\dot{\alpha},b\dot{\beta}}&=C\epsilon^{ab}\epsilon^{\dot{\alpha}\dot{\beta}}\comma\quad
M_{a\dot{\alpha},\beta\dot{b}}=H\epsilon^{a\dot{b}}\epsilon^{\beta\dot{\alpha}}\comma\\
M_{\alpha\dot{a},b\dot{\beta}}&=H\epsilon^{b\dot{a}}\epsilon^{\alpha\dot{\beta}}\comma\quad
M_{\alpha\dot{a},\beta\dot{b}}=C\epsilon^{\dot{a}\dot{b}}\epsilon^{\alpha\beta}\comma
\end{aligned}
\eeq
where
\beq
\begin{aligned}\label{eq:finalwithz}
&
B=-\frac{x^{-}+z^4(x^{+})^3}{ z^2x^{+}(1+x^{+}x^{-})} \comma\,\, E=\frac{z^4x^{+}+(x^{-})^3}{z^2 x^{-}(1+x^{+}x^{-})}\\
 &C=-\frac{i\left[z^4(x^{+})^2-(x^{-})^2\right]}{2z^2\sqrt{x^{+}x^{-}}(1+x^{+}x^{-})}\comma\\
&G=\frac{x^{-}+z^2x^{+}}{2 z\sqrt{x^{+}x^{-}}}\comma\,\, H=\frac{x^{-}-z^2x^{+}}{2 z\sqrt{x^{+}x^{-}}}\period
\end{aligned}
\eeq
There are two allowed solutions depending on the value of $z$, $z=\pm i$.

The bound-state reflection factor $r_a$ is given by
\beq
r_a(u)=\left(\prod_{k=-\frac{a-1}{2}}^{\frac{a-1}{2}}\frac{u+ik+\frac{i}{2}}{u+ik}\right)\frac{\sigma_{B,a}(\bar{u}^{\gamma})}{\sigma_{aa}(u^{\gamma},\bar{u}^{\gamma})}\comma
\eeq
\beq
\sigma_{B,a}(u)\equiv 2^{-E_a(u)}\frac{G(x^{[+a]})}{G(x^{[-a]})}\comma
\eeq
with $f^{[\pm a]}\equiv f(u\pm \frac{i a}{2})$ and
\beq
E_a(u)=\frac{a}{2}+\frac{g}{i}\left(\frac{1}{x^{[-a]}}-\frac{1}{x^{[+a]}}\right)\period
\eeq
\section{Fredholm Determinant in Toy Model}\label{ap:rewriting}
Here we write down an explicit expression for the Fredholm determinant in a simple toy model which only contains a single species of particles without bound states. We also elucidate the relation between the Fredholm determinants and the exact Gaudin determinants \eqref{eq:ratiotoexactG}.

The excited state TBA is given by $\log Y =\phi $ with
\beq\label{eq:apeq1}
\phi (u)\equiv -J \tilde{E}+\sum_{k=1}^{M}\log S (u,u_k) +\log (1+Y)\star K\comma
\eeq
where $S$ is the S-matrix and $K(u,v)\equiv -i\del_u\log S(u,v)$. The rapidities $u_k$'s satisfy the exact Bethe equation
\beq
\phi (u_j)=2\pi i \left(n_j+\frac{1}{2}\right)\comma\qquad n_j \in \mathbb{Z}\period
\eeq
Following the discussion in the main text, one obtains the deformed Fredholm kernel $\hat{G}^{\bullet}\cdot f={\sf S}\cdot f +{\sf I}\cdot f$, with
\beq\label{eq:defSap}
{\sf S}\cdot f\equiv \sum_{k=1}^{M}\frac{i K(u_k,u)}{\del_u \phi(u_k)}f(u_k)\comma\quad {\sf I}\cdot f\equiv \frac{f}{1+1/Y}\star K\period
\eeq

Let us now consider the exact Gaudin norm $\det \left(\del_{u_k}\phi(u_j)\right)$. Using \eqref{eq:apeq1}, we obtain
\beq\label{eq:phaseder}
\del_{u_k}\phi(u_j)=\del_u \phi(u_k)\delta_{jk}-iK(u_k,u_j)+{\sf I}\cdot (\del_{u_k}\epsilon) (u_j)\comma
\eeq
with $\epsilon(u)\equiv \log Y (u)$. To eliminate $\del_{u_k}\epsilon$, we differentiate the excited state TBA,
\beq
\del_{u_k}\epsilon(u)=-iK(u_k,u)+{\sf I}\cdot (\del_{u_k}\epsilon)(u)\comma
\eeq
and invert it to get
\beq
\del_{u_k}\epsilon (u)=\frac{1}{1-{\sf I}}\cdot \left[-iK(u_k,u)\right]\period
\eeq
Plugging it back to \eqref{eq:phaseder}, we obtain
\beq\label{eq:pluggingback}
\del_{u_k}\phi(u_j)=\del_u \phi (u_k)\delta_{jk}-\left.\frac{1}{1-{\sf I}}\cdot \left[iK(u_k,u)\right]\right|_{u=u_j}\period
\eeq

Comparing \eqref{eq:defSap} and \eqref{eq:pluggingback}, we find
\beq
\det (\del_{u_k}\phi(u_j))=\left(\prod_{k=1}^{M}\del_{u}\phi (u_k)\right){\rm Det} (1-\tilde{\sf S})\comma
\eeq
with $\tilde{\sf S}\equiv {\sf S}/(1-{\sf I})$. This relation allows us to relate the Fredholm determinant and the exact Gaudin determinant. The derivation can be readily generalized to $\hat{G}_{+}^{\bullet}$ and to theories with bound states such as $\mathcal{N}=4$ SYM, leading to the identity \eqref{eq:ratiotoexactG}.
\end{document}